\documentclass[conference]{IEEEtran}
\IEEEoverridecommandlockouts
\usepackage{cite}
\usepackage{amsmath,amssymb,amsfonts}
\usepackage{algorithm}
\usepackage{algpseudocodex}
\usepackage{graphicx}
\usepackage{textcomp}
\usepackage[dvipsnames]{xcolor} 
\usepackage{float}
\usepackage{subcaption}
\usepackage{booktabs}
\usepackage{amsmath}
\usepackage{multirow}
\usepackage{colortbl}

\definecolor{gold}{HTML}{FFD700}
\definecolor{silver}{HTML}{C0C0C0}
\definecolor{bronze}{HTML}{CD7F32}
\usepackage{tikz}
\usetikzlibrary{positioning}
\newcommand{\circnum}[1]{{\color{red!75!black}
\tikz[baseline=(char.base)]{
\node[shape=circle, draw, inner sep=1.2pt, minimum size=1.2em, line width=0.9pt] (char) {\footnotesize\textbf{#1}};
}}%
}
\usepackage{booktabs}
\usepackage{enumitem}
\usepackage{environ}
\NewEnviron{observation}[1][]{%
  \par\noindent\fbox{%
    \parbox{\dimexpr\linewidth-2\fboxsep-2\fboxrule}{%
      \if\relax\detokenize{#1}\relax\else\textbf{\textit{#1:}}\quad\fi
      \ignorespaces\BODY
    }%
  }%
  \par
}
\usepackage{hyperref}
\usepackage[para]{footmisc}
\usepackage{pifont}
\usepackage{listings}
\def\BibTeX{{\rm B\kern-.05em{\sc i\kern-.025em b}\kern-.08em
    T\kern-.1667em\lower.7ex\hbox{E}\kern-.125emX}}
\begin{document}

\title{Decoding the Skew: Distribution-Aware MoE Inference with Adaptive Kernel Dispatch\\
\thanks{*This work were conducted during an internship at NVIDIA.}
}

\author{
\IEEEauthorblockN{En-Ming Huang$^{1,2*}$, An-Cheng Chang$^{2}$, Bai-Cheng Jeng$^{2}$, Shih-Hao Hung$^{1}$, H.T. Kung$^{3}$}
\IEEEauthorblockA{$^{1}$Dept. of Computer Science and Information Engineering, National Taiwan University, Taipei, Taiwan}
\IEEEauthorblockA{$^{2}$NVIDIA Corporation, Taipei, Taiwan}
\IEEEauthorblockA{$^{3}$Dept. of Computer Science, Harvard University, Cambridge, USA}
\IEEEauthorblockA{r13922078@csie.ntu.edu.tw, hungsh@csie.ntu.edu.tw, kung@harvard.edu}
}
 
\maketitle

\begin{abstract}
Mixture-of-Experts (MoE) inference consists of sparse expert GEMMs whose shapes vary with the runtime routing distribution. Existing serving systems typically select fused-MoE kernels using static token-count buckets, ignoring the per-expert routing distribution that determines tile padding, memory reuse, and kernel efficiency.
We introduce a distribution-aware framework for modeling and benchmarking MoE inference. The framework combines the compact \emph{Effective Experts} metric with a Dirichlet-based reverse-modeling procedure that generates controllable routing distributions for systematic hardware studies. Using it, we show that the best fused-MoE kernel changes with routing skew and token count. We further present DA-MoE, a GPU-resident kernel-dispatch runtime for NVIDIA GPUs that matches the live routing histogram to offline-tuned distributions and selects a near-optimal fused-MoE kernel without CPU--GPU synchronization. On HumanEval-X serving traces, DA-MoE improves geomean fused-MoE latency by 1.16$\times$ on DeepSeek-V3 and 1.29$\times$ on Kimi~K2, with peak speedups of 1.40$\times$ and 1.56$\times$.
Our implementation is built on top of FlashInfer, and is merged into the main branch through this pull request: \url{https://github.com/flashinfer-ai/flashinfer/pull/4106}.
\end{abstract}

\begin{IEEEkeywords}
    Large Language Models, Mixture-of-Experts, GPU Inference, Tensor Cores
\end{IEEEkeywords}
\section{Introduction}
\label{sec:intro}

\begin{figure}[t]
    \centering
    \includegraphics[width=0.95\linewidth]{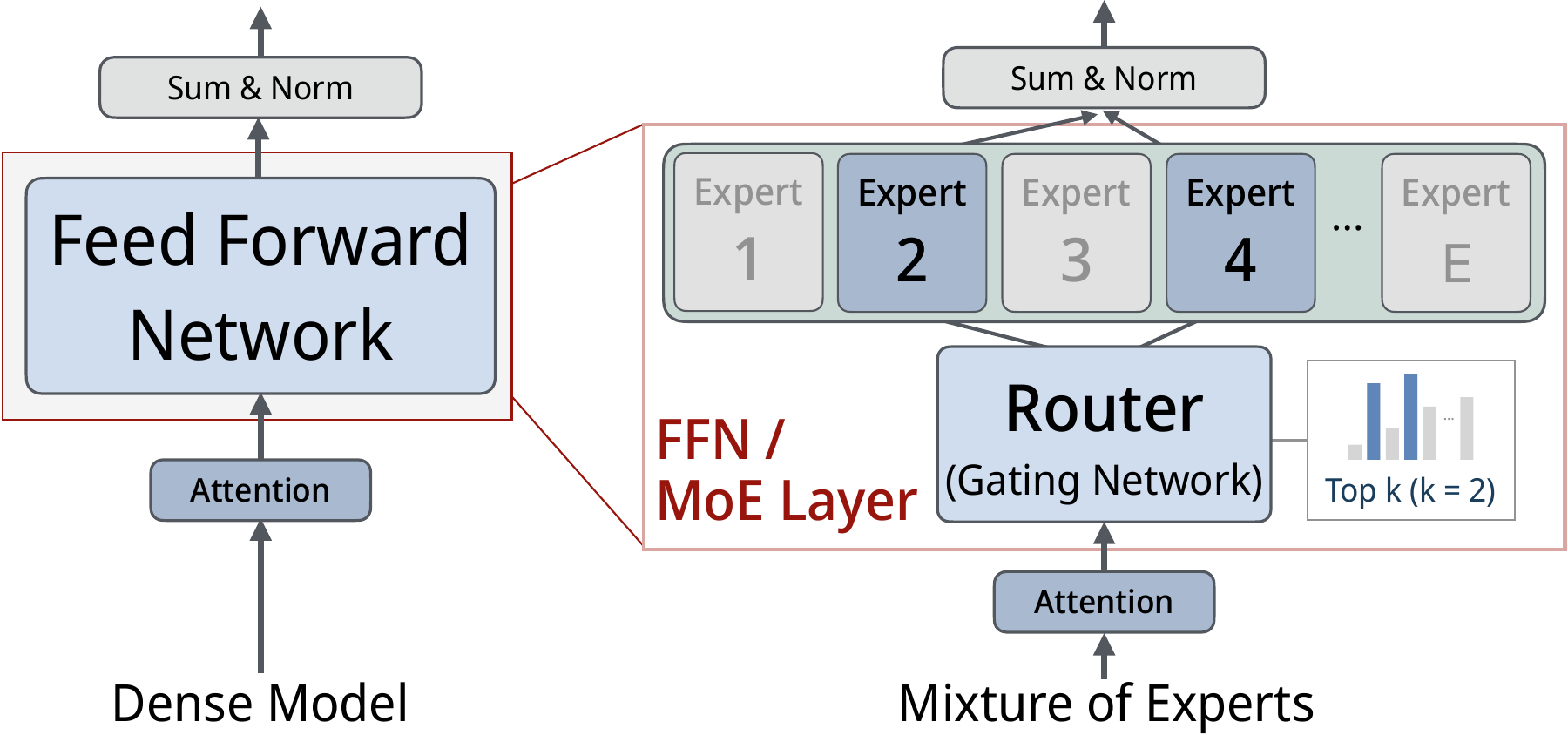}
    \caption{Comparison of dense and MoE layer architectures.}
    \label{fig:moevsdense}
    \vspace{-1em}
\end{figure}

Mixture-of-Experts (MoE) architectures~\cite{jiang2024mixtralexperts,switchtransformer,deepseekai2025deepseekv32pushingfrontieropen,kimiteam2026kimik2openagentic} have become the dominant approach for scaling large language models (LLMs) to trillion-parameter scale, decoupling total model capacity from per-token compute cost. As shown in Fig.~\ref{fig:moevsdense}, by activating only a small subset of expert sub-networks per token, modern MoE models such as DeepSeek-V3~\cite{deepseekai2025deepseekv32pushingfrontieropen} (671B total / 37B activated) keep inference latency tractable while their total capacity grows beyond prior dense models~\cite{brown2020gpt3}. However, this dynamic sparsity produces highly variable workloads across experts and introduces challenges in memory access efficiency, compute utilization, and computation--communication overlap.

These challenges are particularly acute in the small general matrix multiplications (GEMMs) executed by individual experts, where routing imbalance directly determines the shape and efficiency of each computation. At the micro-architectural level, the key lever for managing this variability is the matrix-multiply-accumulate (MMA) tile shape, which determines the tradeoff between data reuse and padding overhead. Yet the right tile depends on a moving target, as MoE routing is not only sparse but also non-uniform across experts, causing per-expert token counts to vary widely across layers and inputs. When these uneven counts are mapped onto fixed-size MMA tiles (e.g., Tensor Cores or systolic arrays~\cite{systolicarray}), two competing inefficiencies emerge. Larger tiles (e.g., up to $256\times256$ on both Blackwell-class NVFP4 Tensor Cores and Google's TPU) maximize data reuse but force under-utilized expert GEMMs to round up, wasting compute cycles on padding trailing columns with zeros; smaller tiles eliminate padding but reload matrix weights more frequently, multiplying memory traffic.

NVIDIA's TensorRT~LLM (TRTLLM)~\cite{trtllm}, the state-of-the-art (SoTA) LLM inference kernel library, responds with kernel specialization, instantiating hundreds of GEMM kernel variants that span different MMA tile shapes (to mitigate tile-padding overhead and memory reuse efficiency) and software-pipelining configurations (to amortize latency). Yet the existing LLM serving frameworks~\cite{trtllm,ye2025flashinfer,kwon2023vllm} uses randomly generated routing assignments to select only ``the best kernel'' for each token count (batch size), ignoring the routing distribution that actually determines per-expert token counts. On the analysis side, recent MoE characterization studies~\cite{yu2026patternschaos,megablocks} rely on statically collected traces and lack mathematical indicators for skew quantification. No existing work provides a generative model of expert skew whose parameters can be tuned for sensitivity analysis in hardware performance models and architectural simulators~\cite{2024llmcompass,reallm2025}.

We observe that closing these gaps requires two coupled pieces: a sweepable model of routing skew and a low-overhead dynamic runtime. We first introduce the \emph{Effective Experts}  metric paired with a Dirichlet distribution model whose concentration parameter $\alpha$ is fitted via binary search. The resulting distributions allow architects to sweep $\alpha$ to generate arbitrary skew levels for controlled simulation and hardware studies. Using this framework, we expose a wide performance variance across TRTLLM's  kernels and quantify the tile-padding overhead of MoE computations on B200's NVFP4 Tensor Cores. 
Second, we present \textbf{DA-MoE}, a distribution-aware runtime that selects near-optimal GPU kernels on a per-layer basis by aligning tile shapes and software configurations with live routing distributions. By leveraging conditional CUDA Graphs, DA-MoE executes kernel selection with negligible overhead while entirely eliminating CPU--GPU synchronization.

The main contributions in this work are fourfold:
\begin{enumerate}[noitemsep,topsep=0pt,leftmargin=*]
    \item \textbf{Mathematical Reverse-Modeling of MoE Routing Skew}: We propose the \emph{Effective Experts} metric and a Dirichlet-based reverse-modeling method, producing parameterized routing distributions that approximate measured real-trace routing and enable controlled studies on real hardware.
    \item \textbf{Quantitative Characterization on NVFP4 Tensor Cores}: Using this framework, we quantify the tile-padding overhead and micro-architectural latency bounds of MoE GEMMs on Blackwell-class NVFP4 Tensor Cores, exposing the performance variance across TRTLLM's generated kernels and the gap to an oracle upper bound.
    \item \textbf{DA-MoE Runtime}: We build DA-MoE for NVIDIA GPUs, which selects a near-optimal tactic per MoE layer invocation based on the live routing distribution using conditional CUDA Graph. On HumanEval-X serving traces, DA-MoE improves geomean fused-MoE latency by 1.16$\times$ on DeepSeek-V3 and 1.29$\times$ on Kimi~K2, with peak speedups of 1.40$\times$ and 1.56$\times$. End-to-end MoE-layer latency improves by 1.15$\times$ and 1.26$\times$ geomean, peaking at 1.37$\times$ and 1.51$\times$, respectively, over SoTA baselines.
    \item \textbf{Co-Design Across SIMT GPUs and Spatial ASICs}: We derive two accelerator-design implications from DA-MoE: SIMT GPUs need fast GPU-resident launch primitives for post-routing kernel selection, while spatial ASICs need routing-aware tile or array partitioning to serve hot experts efficiently without overpadding cold experts.
\end{enumerate}
Our implementation is built on top of FlashInfer, and is merged into the main branch through this pull request: \url{https://github.com/flashinfer-ai/flashinfer/pull/4106}.

The remainder of this paper reviews the necessary background (Section~\ref{sec:background}), develops the routing-skew model and kernel characterization (Section~\ref{sec::modeling_and_bench}), presents DA-MoE (Section~\ref{sec:method}), evaluates its performance (Section~\ref{sec:evaluation}), discusses architectural implications (Section~\ref{sec:discussion}), and concludes with related work and final remarks (Sections~\ref{sec:related}--\ref{sec:conclusion}).

\section{Background}
\label{sec:background}

This section introduces the background for the rest of the paper: modern MoE model structures (\S\ref{sec:bg_moe}), the NVIDIA GPU architecture (\S\ref{sec:bg_gpu}), Tensor Core tile shapes and their effect on MoE kernel execution (\S\ref{sec:bg_tc}), and the Dirichlet distribution underlying our reverse-modeling framework (\S\ref{sec:bg_dirichlet}).

\subsection{Mixture-of-Experts Models}
\label{sec:bg_moe}

As illustrated in Fig.~\ref{fig:moevsdense}, a standard Transformer block alternates an attention layer with a feed-forward network (FFN). In a dense model, the full FFN weight matrix is applied to every token. An MoE model replaces the FFN with $E$ parallel expert sub-networks and a learned router that, per token, selects the top-$k$ experts ($k \ll E$) to activate, decoupling total model capacity from per-token compute cost.

Recent open-source models illustrate the trend toward larger $E$. The largest SoTA open-source dense model is Qwen3-32B~\cite{yang2025qwen3technicalreport} (32B parameters), while leading open-source MoE models have scaled to hundreds of experts and trillions of parameters: DeepSeek-V3~\cite{deepseekai2025deepseekv32pushingfrontieropen} (671B total, top-8 of 256 experts, 37B activated per token), Kimi K2~\cite{kimiteam2026kimik2openagentic} (1T total, top-8 of 384 experts, 32B activated per token), and Llama 4 Maverick~\cite{2025metallama4} (400B total, top-1 of 128 experts). As $E$ grows, per-expert token counts within a batch become smaller and more non-uniform, intensifying the inefficiencies described in Section~\ref{sec:intro}.

\begin{figure}[t]
    \centering
    \includegraphics[width=0.95\linewidth]{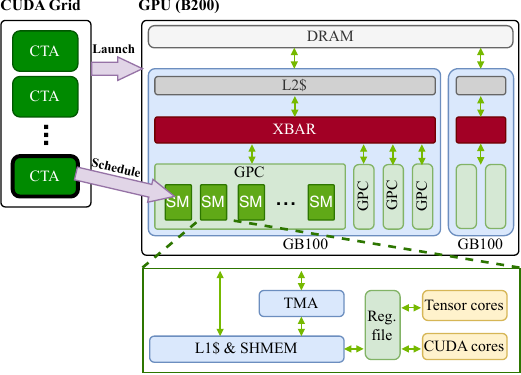}
    \caption{NVIDIA GPU (B200) microarchitecture.}
    \label{fig:gpuarch}
    \vspace{-1em}
\end{figure}

\subsection{NVIDIA GPU Microarchitecture}
\label{sec:bg_gpu}

\textbf{Hardware.}
As illustrated in Fig.~\ref{fig:gpuarch}, an NVIDIA GPU's memory hierarchy consists of two main levels for general data access, including a configurable shared memory and L1 data cache per Streaming Multiprocessor (SM), and a chip-wide L2 cache backed by device memory. Within each SM, CUDA cores support general-purpose single-instruction, multiple-threads (SIMT) execution, while Tensor Cores support MMA operations. Data reaches the compute units through standard load/store instructions issued by the CUDA core pipeline or through the Tensor Memory Accelerator (TMA) with both data paths backed by dedicated buffers and queues. TMA uses dedicated hardware to compute addresses for an $N$-dimensional array and load data into shared memory in the background, freeing CUDA threads from address calculation and overlapping data movement with MMA computation.

\textbf{Software.}
A CUDA kernel is the GPU-side function invoked by the CPU program. Each kernel launch creates a grid of Cooperative Thread Arrays (CTAs), and the GPU dispatches each CTA to an SM. In a tiled GEMM kernel, each CTA is responsible for computing one output tile: it iterates over the $K$ dimension (the K-loop), loading one weight tile and one activation tile into shared memory per iteration, performing the MMA, and accumulating the result. The tile shape therefore determines how much shared memory is consumed per CTA, how many K-loop iterations are required, and how well data can be reused within shared memory before the next load, directly governing both memory traffic and SM occupancy.

\begin{figure}[t]
    \centering
    \includegraphics[width=0.95\linewidth]{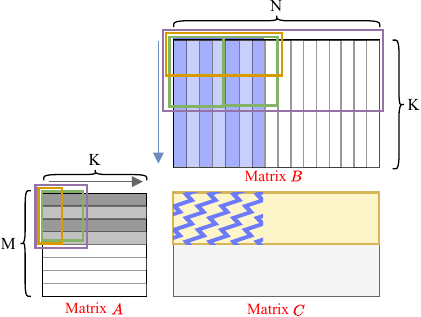}
    \caption{MMA tile shape choices for a single expert GEMM ($C = A \times B$, accumulator omitted for clarity). $A$ is the expert weight matrix ($\mathbb{R}^{M \times K}$); $B$ is the token matrix ($\mathbb{R}^{K \times N}$) where blue columns are real tokens and white columns are zero-padding. {\color{ForestGreen}Green}: small N-tile requires two N-loop iterations, loading $A$ twice. {\color{brown}Brown}: same N-coverage in one pass but with a smaller K-tile (half the K-loop of green). {\color{purple}Purple}: largest N-tile covers all tokens in one pass but incurs greater waste.}
    \label{fig:kernelgen-mmatile}
    \vspace{-1em}
\end{figure}

\subsection{NVIDIA GPU Tensor Core \& TRTLLM Kernel Generation}
\label{sec:bg_tc}

Tensor Cores execute MMA operations over fixed-size tiles of shape $M \times N \times K$. On Blackwell GPUs, the dense NVFP4\footnote{NVFP4 uses E2M1 (2 exponent, 1 mantissa, 1 sign bit) with no Inf/NaN.} Tensor Core instruction family (\texttt{tcgen05.mma}) exposes 80 valid dense tile shapes, ranging from small-$N$ tiles such as $128 \times 8 \times 64$ to large tiles such as $256 \times 256 \times 64$. This range is important for MoE inference as larger tiles reduce the number of MMA instructions and improve data reuse, but can also increase padding when an expert receives only a small number of tokens, wasting computing power. Smaller tiles reduce padding but reload weights more frequently. Selecting optimal tile shapes is a central design variable for MoE models, where the per-expert token count changes across layers and inputs.

Earlier MoE implementations such as Mixtral~8x7B~\cite{jiang2024mixtralexperts} launch one GPU kernel per expert, incurring repeated kernel-launch overhead and leaving SMs idle between launches. NVIDIA's TRTLLM~\cite{trtllm}, the SoTA library for LLM inference, instead uses \emph{fused MoE kernels} that complete all expert computations for an MoE layer within a single launch, amortizing overhead and improving SM utilization. TRTLLM produces hundreds of such kernel variants, each instantiated with a specific combination of MMA tile shape, software-pipelining configuration (e.g., memory pre-fetching depth), and memory load path (e.g., traditional load instructions vs.\ TMA).

The choice of tile shape directly interacts with per-expert token counts. As shown in Fig.~\ref{fig:kernelgen-mmatile}, each expert computes $C = A \times B$ where $A$ is the expert weight matrix and $B$ holds the tokens assigned to that expert (blue columns); the remaining white columns are zero-padded. The three colored tiles show different choices along the N dimension: the green tile requires two N-loop iterations (loading $A$ twice), while the brown and purple tiles each cover all tokens in one pass but differ in padding waste and K-loop tile size. Although the optimal tile shape shifts with the routing distribution, as detailed in Section~\ref{sec::modeling_and_bench}, the per-kernel configurations are closed-source as NVIDIA distributes only pre-compiled binary files, leaving the execution logic opaque to researchers. In this work, we examine how the \textit{Tile-N} dimension affects MoE inference performance, as its optimal value depends directly on the dynamic token routing distribution.

\subsection{Dirichlet Distribution}
\label{sec:bg_dirichlet}

\begin{figure}[t]
    \centering
    \includegraphics[width=0.95\linewidth]{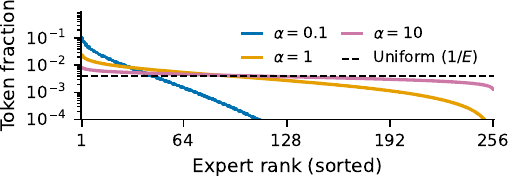}
    \caption{Per-expert token fractions (log-scale)  from $\text{Dir}(\alpha)$ with $E=256$ experts. Small $\alpha$ concentrates tokens on few experts; large $\alpha$ approaches the uniform baseline (dashed).}
    \label{fig:dirichlet-hist}
\end{figure}

An MoE forward pass distributes tokens among experts and induces a routing vector $\mathbf{p} = (p_1, \ldots, p_E)$. Each element $p_i$ denotes the fraction of tokens routed to expert $i$, which gives $p_i \geq 0$ and $\sum_i p_i = 1$. To model variation in this routing vector, we use a Dirichlet distribution, whose samples are valid probability vectors over the $E$ experts. In its general form, $\mathrm{Dir}(\boldsymbol{\alpha})$ is parameterized by a positive vector $\boldsymbol{\alpha} = (\alpha_1, \ldots, \alpha_E)$ and has density
\begin{equation}
f(\mathbf{p};\,\boldsymbol{\alpha}) = \frac{1}{B(\boldsymbol{\alpha})} \prod_{i=1}^{E} p_i^{\,\alpha_i - 1}.
\label{eq:dirichlet}
\end{equation}
Here, $B(\boldsymbol{\alpha})$ is the multivariate Beta function. The density specifies the probability of sampling each routing vector, while $\boldsymbol{\alpha}$ controls the shape of the distribution.

Each $\alpha_i$ acts as a prior count that biases samples toward coordinate $i$. We use the symmetric case, where $\alpha_i = \alpha$ for all $i$, so the scalar $\alpha$ is the only input parameter. A sample $\mathbf{p} \sim \mathrm{Dir}(\alpha,\ldots,\alpha)$ assigns a routing probability to each expert. Fig.~\ref{fig:dirichlet-hist} illustrates the resulting synthetic distributions for different values of $\alpha$. Large values produce samples close to the uniform vector $(1/E, \ldots, 1/E)$, whereas small values concentrate most probability on one or a few experts. The scalar $\alpha$ provides a single control parameter that spans nearly uniform routing ($\alpha \to \infty$) and highly skewed routing ($\alpha \to 0$).

The Dirichlet distribution directly generates valid routing probability vectors over $E$ experts. In contrast, Gaussian, Beta, and exponential distributions do not jointly constrain all expert probabilities to sum to one. In Section~\ref{sec:eval-dist}, we also evaluate these alternatives empirically for workload modeling.

\section{Modeling and Benchmarking}
\label{sec::modeling_and_bench}

This section develops a quantitative framework for studying MoE kernel performance under routing skew. We first introduce the \emph{Effective Experts} metric and a Dirichlet-based reverse-modeling procedure that generates controlled routing distributions (\S\ref{sec::modeling}). We then apply this metric to characterize the layer-wise routing behavior of DeepSeek-V3 (\S\ref{sec:effexp-layer}). Finally, we benchmark TRTLLM's fused MoE kernels under these controlled distributions, revealing that tile shape, pipelining depth, and memory-load path must all be co-adapted to the routing distribution (\S\ref{sec:benchmark}, \S\ref{sec:nosinglebest}).

\subsection{Routing Skew Modeling}
\label{sec::modeling}

In this paper, we propose \emph{Effective Experts} ($eff\_exp$), a metric that directly expresses routing concentration as an interpretable effective number of active experts and is convenient for controlled hardware sweeps. 
\begin{equation}
eff\_exp = \frac{\left(\sum_{i=1}^{E} t_i\right)^2}{\sum_{i=1}^{E} t_i^2},
\label{eq:effexp}
\end{equation}
where $t_i$ is the number of tokens routed to expert $i$ in a given forward pass and $E$ is the total number of experts. Intuitively, $eff\_exp$ estimates how many experts carry the tokens. A high value represents that traffic is spread over many experts, whereas a low value indicates that tokens are concentrated on a smaller expert subset and routing is more skewed. The metric ranges over $[1,\, E]$, equaling $E$ under uniform routing and $1$ when all tokens are directed to a single expert. Unlike \textit{variance}, which varies with total token counts, or \textit{entropy}~\cite{yu2026patternschaos}, which expands logarithmically with expert count ($E$), \textit{effective experts} is scale-invariant across batch sizes and provides an intuitive, physical proxy for workload modeling.

To generate synthetic routing distributions that match a target $eff\_exp^*$, we use a symmetric Dirichlet distribution with concentration parameter $\alpha$ for simplicity. As $eff\_exp$ increases monotonically with $\alpha$, the corresponding $\alpha$ can be derived by binary search. At each iteration, we draw $S$ samples from $\mathrm{Dir}(\alpha_{mid})$, compute their mean $eff\_exp$, and narrow the interval accordingly, as detailed in Algorithm~\ref{alg:binsearch}.

\begin{algorithm}[t]
\caption{Binary Search for Dirichlet $\alpha$}
\label{alg:binsearch}
\begin{algorithmic}[1]
\Require Target $eff\_exp^*$, number of experts $E$, samples $S$, iterations $T$
\Require Search range $[\alpha_{\min}, \alpha_{\max}] = [10^{-6}, 10^{6}]$
\Ensure $\alpha$ such that sampled $eff\_exp \approx eff\_exp^*$
\State $lo \gets \alpha_{\min}$, $\quad hi \gets \alpha_{\max}$
\For{$t = 1$ \textbf{to} $T$}
    \State $\alpha_{mid} \gets (lo + hi) / 2$
    \State Draw $\mathbf{p}^{(1)}, \ldots, \mathbf{p}^{(S)} \sim \mathrm{Dir}(\alpha_{mid}, \ldots, \alpha_{mid})$
    \State $v \gets \frac{1}{S} \displaystyle\sum_{s=1}^{S} eff\_exp\!\left(\mathbf{p}^{(s)}\right)$
    \If{$v < eff\_exp^*$}
        \State $lo \gets \alpha_{mid}$ \Comment{too skewed; increase $\alpha$}
    \Else
        \State $hi \gets \alpha_{mid}$ \Comment{too uniform; decrease $\alpha$}
    \EndIf
\EndFor
\State \Return $\alpha_{mid}$
\end{algorithmic}
\end{algorithm}

\begin{figure}[t]
    \centering
    \includegraphics[width=\linewidth]{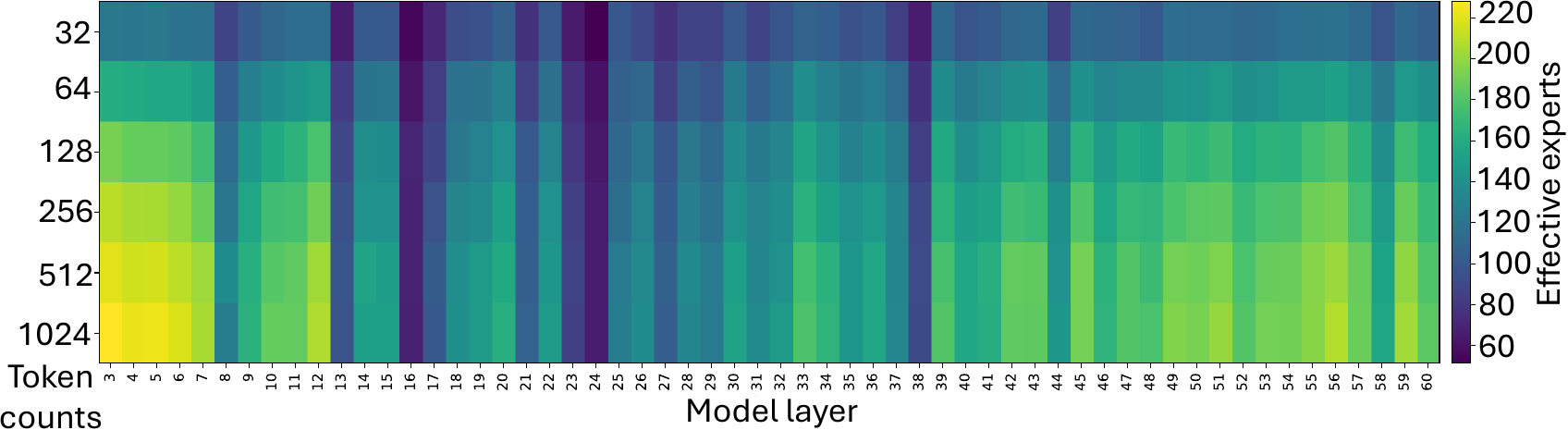}
    \caption{$eff\_exp$ across model layers and token counts for DeepSeek-V3 (NVFP4) on the MMLU dataset~\cite{hendrycks2021mmlu}. Middle layers exhibit consistently low $eff\_exp$ (concentrated routing), while early and late layers are more uniform.}
    \label{fig:effexp-layer}
\end{figure}

\begin{figure*}[t]
    \hspace*{-1.3em}
    \includegraphics[width=1.1\textwidth]{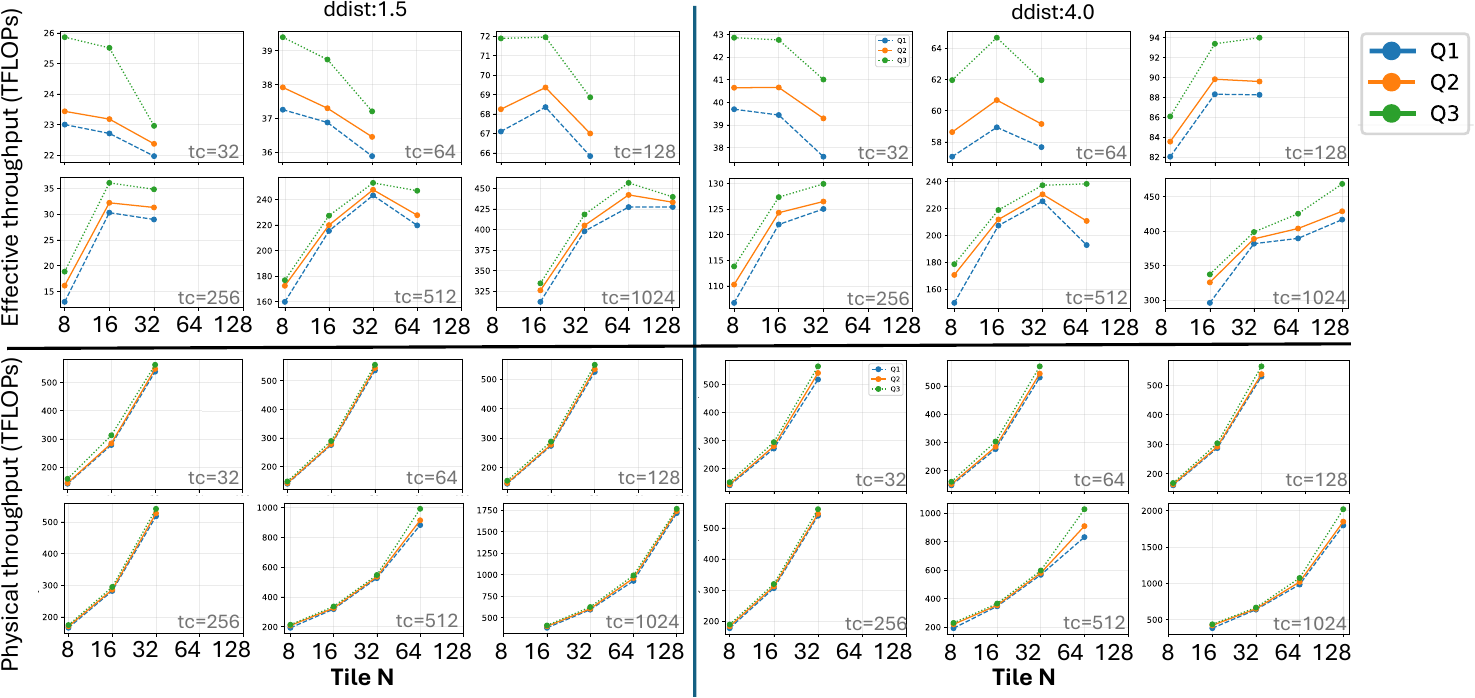}
    \caption{Effective throughput (top rows) and physical throughput (bottom rows) of TRTLLM MoE kernels as a function of \textit{Tile-N}, under \texttt{ddist:1.5} (left) and \texttt{ddist:4.0} (right) for token counts 32--1024. Each point shows Q1/Q2/Q3 percentiles across all kernels sharing the same \textit{Tile-N}. Physical throughput increases monotonically with \textit{Tile-N} and Q1--Q3 bands are tight. Effective throughput is non-monotonic: for small batches, large \textit{Tile-N} incurs excessive zero-padding, degrading useful throughput; for larger batches, an intermediate \textit{Tile-N} is optimal. The Q1--Q3 spread in effective throughput exceeds 10\% at the same \textit{Tile-N}, reflecting the impact of software-pipelining and memory-path configuration. (TC = token counts)}
    \label{fig:ddist-difftileN}
    \vspace{-1em}
\end{figure*}

\begin{figure}[t]
    \centering
    \includegraphics[width=.99\linewidth]{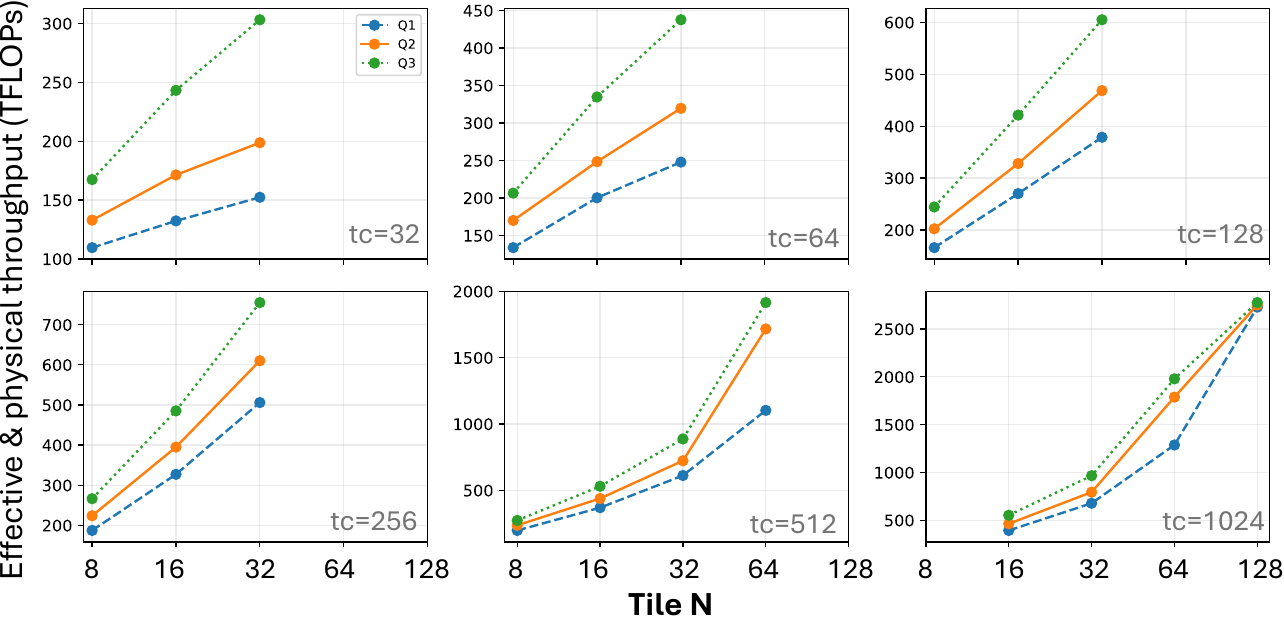}
    \caption{Effective and physical throughput when all tokens are routed to a single expert (dense-model behavior). Performance increases monotonically with \textit{Tile-N} across all token counts, as larger tiles maximize weight-matrix data reuse.}
    \label{fig:effthru-single}
\end{figure}

We denote the resulting parameterization \texttt{ddist:X}, a $\mathrm{Dir}(\alpha)$ distribution targeting $eff\_exp^* = E / X$, with the parameter $X$ acting as a skew multiplier. Increasing $X$ reduces the target number of effectively used experts and concentrates tokens on a smaller subset of experts. At $X = 1$, $eff\_exp^* = E$, so routing is near-uniform across all experts; at $X = E$, $eff\_exp^* = 1$, so the distribution approaches the maximally skewed case in which one expert dominates. This notation decouples routing skew from model-specific expert counts and provides a stable axis for sweeping distributions.

\subsection{Layer-Wise Routing Skew in DeepSeek-V3}
\label{sec:effexp-layer}

Fig.~\ref{fig:effexp-layer} reports $eff\_exp$ measured layer-by-layer for DeepSeek-V3 on the MMLU dataset~\cite{hendrycks2021mmlu}. It can be observed that the routing skew is highly layer-dependent as more central layers exhibit persistently low $eff\_exp$ (concentrated routing), while early layers (3--12) and late layers (45--60) remain more uniform. The layer-wise distribution shifts substantially in all token counts, from values around 80 in the most skewed layers to values around 180 or higher in the more uniform regime. This heterogeneity implies that a single, statically chosen kernel configuration will be mismatched to a substantial fraction of layers at any given token count.

\begin{observation}[Observation 1]
MoE routing skew depends on layers and token counts. In our DeepSeek-V3 trace, $eff\_exp$ varies by over $2\times$ across layers at each token count, so a dispatch policy should adapt to the observed routing distribution rather than assume one fixed skew level for the whole model.
\end{observation}

\subsection{Kernel Performance Under Controlled Distributions}
\label{sec:benchmark}

\subsubsection{Throughput Definitions}

We profile TRTLLM's fused MoE kernels under controlled \texttt{ddist:X} distributions and report two throughput metrics: (a) \textbf{Physical throughput}: total MMA FLOPs issued per second, including both useful and zero-padded compute;
(b) \textbf{Effective throughput}: physical throughput scaled by the fraction of non-padded tokens, reflecting only computation that contributes to valid outputs.

\subsubsection{Physical vs.\ Effective Throughput}

Fig.~\ref{fig:ddist-difftileN} reports both metrics for \texttt{ddist:1.5} and \texttt{ddist:4.0} across \textit{Tile-N} $\in \{8, 16, 32, 64, 128\}$ and token counts 32--1024. Physical throughput (bottom rows) increases monotonically with \textit{Tile-N} regardless of token count or distribution, since larger tiles reuse the expert weight matrix over more token columns, raising arithmetic intensity toward the compute roofline.

On the other hand, since effective throughput (top rows) includes the penalty from zero-padded work, its trend is non-monotonic. The preferred \textit{Tile-N} follows the number of tokens assigned to each active expert. Under \texttt{ddist:1.5}, small batches favor the smallest tiles, while larger batches move the optimum toward larger \textit{Tile-N} values; the preferred tile grows from \textit{Tile-N}=8 at token counts 32--64 to \textit{Tile-N}=64 at token count 1024. This progression reflects the tradeoff between zero-padding and memory reuse. Small tiles reduce compute waste when each expert receives few tokens, while larger tiles improve expert weight reuse by increasing arithmetic intensity. Higher skew shifts the same tradeoff toward larger tiles. Under \texttt{ddist:4.0}, tokens concentrate on fewer experts, increasing the token counts of hot experts and moving the preferred \textit{Tile-N} one step higher in the tile sequence. The Q1--Q3 spread in effective throughput in both distributions exceeds 10\% at the same \textit{Tile-N}, indicating that the configurations of software-pipelining depth and memory-load path could affect the actual kernel performance and thus cannot be overlooked.

\subsubsection{Single-Expert (Dense Model) Routing}

Fig.~\ref{fig:effthru-single} shows the limiting case where all tokens route to a single expert, equivalent to dense-model behavior. As zero-padding is absent, effective throughput equals physical throughput, so performance increases monotonically with \textit{Tile-N} across all token counts. This baseline confirms that larger tiles are unconditionally better only when routing is perfectly concentrated. In contrast, when the routing distribution becomes non-uniform, the optimal \textit{Tile-N} shifts downward because larger tiles waste more computation on padding.

\vspace{0.1em}
\begin{observation}[Observation 2]
The optimal \textit{Tile-N} depends on both the routing distribution and the token count. Under skewed routing, smaller tiles reduce padding waste; under uniform routing or large batches, larger tiles improve data reuse. Tile shape selection cannot be decoupled from routing skew.
\end{observation}

\subsection{No Single Kernel Dominates Across Distributions}
\label{sec:nosinglebest}

\begin{figure}[t]
    \centering
    \includegraphics[width=\linewidth]{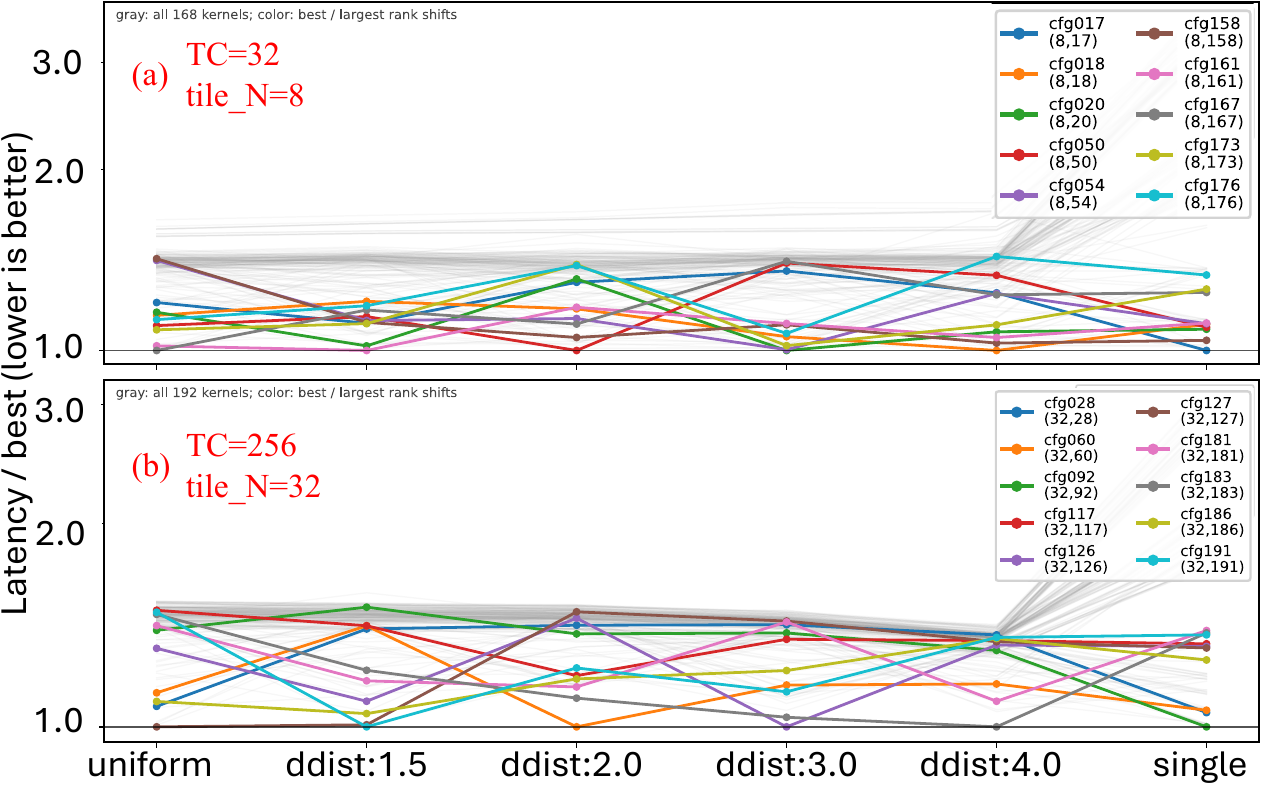}
    \caption{Normalized latency and the top-performing kernels (colored) as routing distribution is swept from uniform through \texttt{ddist:1.5}--\texttt{4.0} to single-expert. (a)~TC\,=\,32, \textit{Tile-N}\,=\,8 (168 kernels). (b)~TC\,=\,256, \textit{Tile-N}\,=\,32 (192 kernels). The best kernel changes across distributions in both settings, confirming that no single kernel dominates all routing patterns.}
    \label{fig:no-single-best}
\end{figure}

The preceding analysis shows that the optimal tile shape depends on both routing distribution and token count. Beyond tile shape, each \textit{Tile-N} still admits multiple kernel configurations that differ in pipelining configuration and memory-load path. We next examine whether any single configuration remains optimal across distributions at a fixed \textit{Tile-N}.

Fig.~\ref{fig:no-single-best} sweeps routing from uniform through \texttt{ddist:1.5}, \texttt{2.0}, \texttt{3.0}, \texttt{4.0} to single-expert and plots the normalized latency (latency\,/\,best-in-distribution) for all TRTLLM kernels sharing the same \textit{Tile-N}. Gray lines show all kernels, while colored lines highlight the top-performing ones. In both token count of 32 and 256, the kernel configuration that achieves the lowest latency changes across distributions. This result directly motivates the distribution-aware dispatch system described in Section~\ref{sec:method}, since the optimal kernel shifts with the routing distribution observed at runtime.
\vspace{0.3em}
\begin{observation}[Observation 3]
Even at a fixed \textit{Tile-N}, no single kernel configuration dominates all routing distributions. The dispatch system must select among kernel configurations at runtime based on the observed routing distribution.
\end{observation}

\section{DA-MoE: Distribution-Aware Kernel Dispatch}
\label{sec:method}

\begin{figure*}[t]
    \centering
    \includegraphics[width=0.9\textwidth]{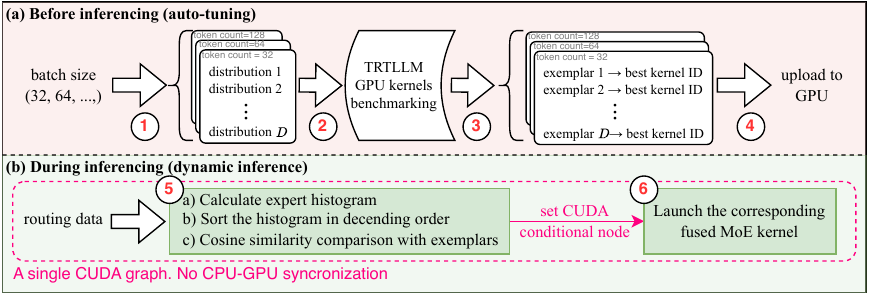}
    \caption{Distribution-aware dispatch flow. \textbf{(a)~Offline auto-tuning}: (1)~for each token-count bucket, generate synthetic routing distributions via \texttt{ddist:X}; (2)~benchmark them against all available TRTLLM kernels; (3)~record the best kernel ID per distribution and form an exemplar--kernel pair, where the exemplar is the sorted, L2-normalized per-expert token vector; (4)~upload the bucket-specific exemplar tables to the GPU. \textbf{(b)~Online dispatch}: (5)~build the per-expert histogram from the router logits, sort it, and match it against the exemplar table for the current token-count bucket by cosine similarity; (6)~launch the selected fused MoE kernel through a conditional CUDA Graph, entirely on the GPU with no CPU--GPU synchronization.}
    \label{fig:damoe-flow}
\end{figure*}

Section~\ref{sec::modeling_and_bench} shows that no single kernel configuration is optimal across routing distributions since the best choice changes with both the layer and the token count. Exploiting this behavior during inference imposes two challenges. First, the system must identify the kernel that best matches the current routing distribution. Second, it must apply this choice without introducing CPU-side dispatch overhead that could offset the performance benefit, because the router network produces the routing distribution on the GPU while conventional kernel invocation is controlled by the CPU. DA-MoE addresses both requirements through an offline auto-tuning stage that maps representative routing distributions to their fastest kernels (\S\ref{sec:autotune}) and an online, fully GPU-resident dispatch path built on conditional CUDA Graphs that applies the resulting selection without CPU intervention (\S\ref{sec:online},~\S\ref{sec:condcg}), as illustrated in Fig.~\ref{fig:damoe-flow}.


At a high level, as depicted in Fig.~\ref{fig:damoe-flow}, our DA-MoE workflow consists of two stages. The offline auto-tuning stage profiles candidate fused MoE kernels and builds a separate lookup table for each token-count bucket, while the online dynamic path first selects the bucket for the current token count, queries that bucket's table, and launches the selected kernel without CPU intervention. Similar profiling-and-dispatch flows already appear in common LLM inference frameworks such as TRTLLM, vLLM and FlashInfer~\cite{trtllm,kwon2023vllm,ye2025flashinfer}; however, these flows typically use random router logits during tuning, choose a single kernel configuration for each token-count bucket, and instantiate one CUDA Graph per bucket (e.g., 32, 64, $\ldots$). DA-MoE instead makes the routing distribution explicit, so each token-count bucket has its own set of exemplars and can contain multiple distribution-specific kernel choices while still preserving graph-based launch efficiency.

\subsection{Offline Auto-Tuning}
\label{sec:autotune}

The offline stage constructs one lookup table per token-count bucket. Each table maps representative routing distributions for that bucket to the fused MoE kernels that execute them with the lowest latency. As shown in Fig.~\ref{fig:damoe-flow}(a), for each token count we \circnum{1} generate synthetic routing distributions by sweeping \texttt{ddist:X} over a range of skew levels, using the Dirichlet reverse-modeling procedure from \S\ref{sec::modeling}. We then \circnum{2} benchmark each distribution against all available TRTLLM fused MoE kernels and \circnum{3} record the kernel ID with the lowest latency for that distribution.

We represent each distribution with an \emph{exemplar} derived from the vector of token counts across experts. The vector is sorted in descending order and L2-normalized to unit norm. Sorting removes dependence on expert identity, so two layers with the same skew profile but different hot experts map to the same representation, while L2 normalization removes residual scale differences within a token-count bucket. If multiple samples ($S$) are given for a distribution, we use the centroid of the samples as the exemplar, and the kernel with the fastest mean performance is selected. The offline stage produces, for each token-count bucket, a dispatch table that maps each exemplar to the kernel~ID selected by profiling. We \circnum{4} upload these tables to GPU memory before inference.

\subsection{Online Dispatch}
\label{sec:online}

At inference time, the routing distribution of an MoE layer is known only after the router executes, so kernel selection lies on the critical path. As shown in Fig.~\ref{fig:damoe-flow}(b), once the router produces per-token expert logits, we \circnum{5} build the per-expert histogram by counting how many tokens are routed to each expert and sorting the histogram in descending order. The dispatcher compares this query with the exemplars in the current token-count bucket using cosine similarity, which is a single dot product per exemplar because the exemplars are unit vectors. The exemplar with the highest score provides the kernel ID for the current layer invocation.

The histogram construction, sorting, similarity search, and dispatch decision all execute on the GPU. This placement is essential as a CPU-side selector would have to copy the routing histogram back to the host. Keeping the decision on the device removes this synchronization from the critical path.

\subsection{GPU-Side Dispatch via Conditional CUDA Graphs}
\label{sec:condcg}

\begin{figure}[t]
    \centering
    \includegraphics[width=\linewidth]{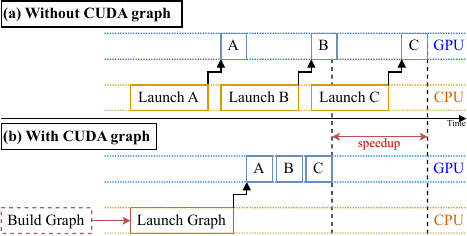}
    \caption{Kernel launch with and without CUDA Graphs. \textbf{(a)~Without CUDA Graphs}, each kernel is launched through the CPU, introducing launch latency. \textbf{(b)~With CUDA Graphs}, the runtime captures the kernel sequence once and replays it as a single graph execution.}
    \label{fig:cudagraph}
\end{figure}

\begin{figure}[t]
    \centering
    \includegraphics[width=\linewidth]{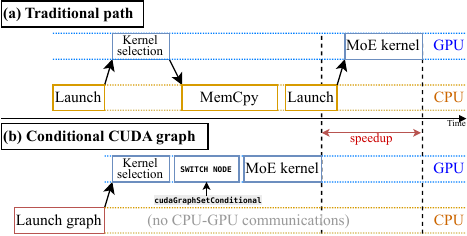}
    \caption{Traditional vs.\ conditional CUDA Graphs. \textbf{(a)~Traditional graphs} require CPU copy-back and subgraph relaunch for data-dependent dispatch. \textbf{(b)~Conditional graphs} use a device-side \texttt{SWITCH} to keep selection and launch on the GPU.}
    \label{fig:cond-cg}
\end{figure}

The remaining obstacle is the launch mechanism for low-latency inference. LLM inference issues many small kernels per model layer, and launching each kernel separately through the CPU exposes per-launch overhead on the critical path. CUDA Graphs mitigate this overhead by capturing a fixed kernel sequence once and replaying it as a single graph execution, thereby amortizing CPU launch cost. As illustrated in Fig.~\ref{fig:cudagraph}, instead of issuing individual CPU launches for each kernel in the decoding path, the runtime replays a pre-captured graph that contains the same kernel sequence. This property is why modern LLM serving frameworks~\cite{kwon2023vllm,trtllm,zheng2024sglang} rely heavily on CUDA Graphs.

A standard CUDA Graph, however, represents a static sequence. It cannot make a data-dependent branch to one of several kernels without leaving the graph. Fig.~\ref{fig:cond-cg}(a) shows the consequence for distribution-aware dispatch under traditional graphs. The graph must be split at the dispatch point, and the selected kernel ID must be returned to the CPU so that the CPU can launch the matching subgraph. This design reintroduces the same CPU--GPU synchronization that our GPU-resident selector is meant to avoid.

We avoid this overhead with conditional CUDA Graphs, introduced in CUDA~12.8. A \texttt{cudaGraphConditional} node embeds a device-side \texttt{SWITCH} branch directly inside the graph. Fig.~\ref{fig:cond-cg}(b) shows our dispatch path: the kernel ID selected in \S\ref{sec:online} drives the switch, and the corresponding fused MoE kernel is launched without returning control to the host. Thus, after the online matcher produces a kernel ID, the graph \circnum{6} launches the selected fused MoE kernel through the device-side conditional branch. Histogram construction, exemplar matching, branch selection, and kernel launch all execute within one graph context, keeping distribution-aware selection GPU-resident and low overhead relative to the kernel execution it controls. Currently, no work has explored the application of conditional CUDA Graphs on LLM  systems.

\section{Evaluation}
\label{sec:evaluation}

This section evaluates the performance of DA-MoE in realistic LLM serving. The central concern is whether a runtime selector can improve over the static per-bucket kernel choice used by production MoE backends, while preserving the CUDA Graph execution model needed for low-latency decoding. We compare DA-MoE against \emph{NoDA}, which uses random inputs to profile and select the best kernel for each token-count, as in common auto-tuning flows used by TRTLLM~\cite{trtllm} and FlashInfer~\cite{ye2025flashinfer}. We also report an \emph{oracle} that selects the fastest available kernel for each individual invocation.

We report two latency metrics. \emph{Fused-MoE latency} measures only the expert-computation kernel and isolates the benefit of distribution-aware kernel selection, while \emph{end-to-end MoE-layer latency} measures the full CUDA Graph path, including router selection, routing-metadata construction, DA-MoE dispatch, and fused-MoE execution. The first metric shows how much opportunity exists in the kernel choice itself; the second shows how much of that opportunity remains after the GPU-resident selection cost is included.

The evaluation proceeds as follows.
We first establish the headline result on real workloads (\S\ref{sec:eval-real}).
We then quantify the cost of keeping the dispatch decision on GPU through conditional CUDA Graphs (\S\ref{sec:eval-condcg}).
Next, we validate the assumption that synthetic Dirichlet distributions match real routings (\S\ref{sec:eval-dist}).
Finally, we study how the gain scales with routing skew (\S\ref{sec:eval-ablation-imbalance}) and expert-parallelism degree (\S\ref{sec:eval-ablation-ep}).

\subsection{Experimental Setup}
\label{sec:eval-setup}

All experiments run on NVIDIA B200 GPUs connected with NVLink under CUDA~13.1. We implement DA-MoE on FlashInfer v0.6.6~\cite{ye2025flashinfer} and integrate it with vLLM~\cite{kwon2023vllm} for LLM serving. Experiments are evaluated on two large MoE-based LLMs: DeepSeek-V3~\cite{deepseekai2025deepseekv32pushingfrontieropen} running on four GPUs with expert parallelism (EP) EP=4, and Kimi~K2~\cite{kimiteam2026kimik2openagentic} on six GPUs with EP=6. The models specification are listed in Table~\ref{tab:s5-model-specs}.

\begin{table}[t]
  \caption{Model specifications.}
  \label{tab:s5-model-specs}
  \centering
  \begin{tabular}{lrr}
    \toprule
    \textbf{Specification} & \textbf{DeepSeek-V3} & \textbf{Kimi K2} \\
    \midrule
    Total parameters & 671B & 1T \\
    Activated parameters & 37B & 32B \\
    Number of experts & 256 & 384 \\
    Experts selected per token & 8 & 8 \\
    Model size (NVFP4) & 415GB & 590GB \\
    Quantization & NVFP4 & NVFP4 \\
    \bottomrule
  \end{tabular}
\end{table}

We use two workloads for distinct purposes. MMLU~\cite{hendrycks2021mmlu} is used only for routing-distribution analysis, matching Section~\ref{sec:benchmark}, while HumanEval-X~\cite{codegeex2023} provides the real-world serving workload, testing DA-MoE on a dataset different from the one used for workload analysis. Decoding uses temperature 0.3 with each configuration executed eight times, and we report the median to suppress noise.

The offline auto-tuning sweep covers token-count buckets $\{32,64,128,256,512\}$ and Dirichlet skew settings $\{1.1,1.3,1.5,1.7,2.0,2.5,4.0\}$ with number of samples \(S\) of 10 for each skew, covering the range of skews observed in Section~\ref{sec::modeling}. We focus on token counts 32--512, where low-latency serving is most demanding as tile padding, execution latency and memory efficiency are most visible.

\subsection{Real-Workload MoE Latency}
\label{sec:eval-real}

\begin{figure}[t]
    \centering
    \includegraphics[width=0.95\linewidth]{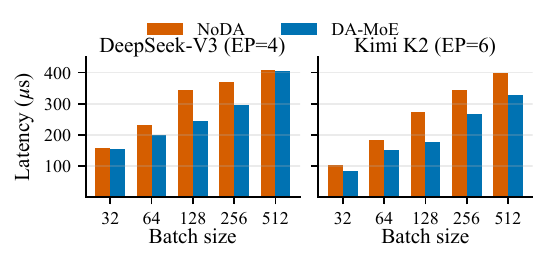}
    \caption{Fused-MoE latency for DeepSeek-V3 and Kimi~K2 across token counts 32--512. DA-MoE improves geomean latency by 1.16$\times$ and 1.29$\times$ over NoDA, respectively.}
    \label{fig:s5-primary-fc-bar}
    
\end{figure}

\begin{figure}[t]
    \centering
    \includegraphics[width=\linewidth]{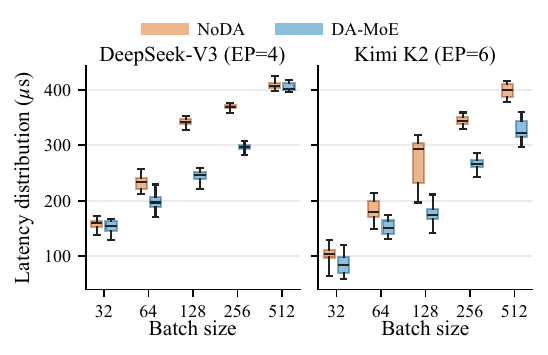}
    \caption{Per-invocation fused-MoE latency distribution for NoDA and DA-MoE, shown as min, Q1, median, Q3, and max across the captured MoE-layer invocations.}
    \label{fig:s5-primary-fc-box}
    
\end{figure}

Fig.~\ref{fig:s5-primary-fc-bar} presents the geometric mean fused-MoE latency on the real-world workload.
On DeepSeek-V3, DA-MoE improves geomean fused-MoE latency by 1.16$\times$ over NoDA, with a peak speedup of 1.40$\times$.
The gains are largest at token counts 64--256, where routing skew and token count jointly expose alternative kernels that are substantially faster than the static bucket choice.
Kimi~K2 shows a larger geomean improvement of 1.29$\times$, peaking at 1.56$\times$ at token count~128.
The stronger gain on Kimi~K2 follows from its larger expert pool, so routing imbalance is more pronounced, giving distribution-aware selection more room to improve.
Fig.~\ref{fig:s5-primary-fc-box} further reports the latency distribution for each MoE layer and decoding step. DA-MoE shifts the latency in all metrics downward, indicating that the gain comes from consistently better kernel selection rather than from a few outliers.
\begin{table}[t]
    \centering
    \caption{Fused-MoE latency normalized to the oracle.}
    \label{tab:s5-fc-oracle}
    \setlength{\tabcolsep}{3.0pt}
    \begin{tabular}{llrrrrr}
        \toprule
        \multirow{2}{*}{Model} & \multirow{2}{*}{Method} & \multicolumn{5}{c}{Token count} \\
        \cmidrule(lr){3-7}
        & & 32 & 64 & 128 & 256 & 512 \\
        \midrule
        \multirow{3}{*}{DS-V3}
        & NoDA & 103.8\% & 118.8\% & 140.1\% & 125.9\% & 121.5\% \\
        & DA-MoE & 100.3\% & 101.1\% & 100.3\% & 101.0\% & 120.6\% \\
        & Oracle & 100.0\% & 100.0\% & 100.0\% & 100.0\% & 100.0\% \\
        \midrule
        \multirow{3}{*}{Kimi K2}
        & NoDA & 122.7\% & 123.0\% & 158.5\% & 131.3\% & 126.7\% \\
        & DA-MoE & 102.6\% & 102.5\% & 101.8\% & 101.6\% & 103.7\% \\
        & Oracle & 100.0\% & 100.0\% & 100.0\% & 100.0\% & 100.0\% \\
        \bottomrule
    \end{tabular}
\end{table}

To evaluate DA-MoE's ability to select effective kernel configurations, Table~\ref{tab:s5-fc-oracle} quantifies how closely each method approaches an oracle that selects the fastest available kernel for each individual invocation.
On DeepSeek-V3, DA-MoE remains within 1.1\% of the oracle for token counts 32--256, whereas NoDA is 26--40\% slower over the same range.
At token count 512, both DA-MoE and NoDA sit approximately 20\% above the oracle, indicating that the oracle selects a kernel outside the coverage of both the exemplar matcher and the static bucket choice.
For Kimi~K2, DA-MoE stays within 4\% of the oracle across all evaluated token counts, while NoDA trails the oracle by up to 59\%.

\subsection{Overhead of Conditional CUDA Graph Dispatch}
\label{sec:eval-condcg}

\begin{table}[t]
    \centering
    \caption{Latency of a data-dependent dispatch microbenchmark: a producer kernel computes a value that selects which of two follow-up kernels to launch.}
    \label{tab:s5-condcg}
    \setlength{\tabcolsep}{4pt}
    \begin{tabular}{lcc}
        \toprule
        Dispatch path & Latency ($\mu$s) & Branch overhead ($\mu$s) \\
        \midrule
        Copy to host, branch on CPU      & 19.1 & 15.0 \\
        Conditional CUDA Graph (on GPU)  & 12.3 & \ 8.2 \\
        Fixed graph, no branch           & \ 4.1  & ---  \\
        \bottomrule
    \end{tabular}
\end{table}

\begin{figure}[t]
    \centering
    \includegraphics[width=\linewidth]{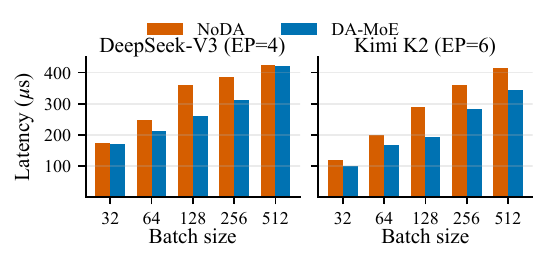}
    \caption{End-to-end MoE-layer latency (the full captured CUDA Graph: router selection, routing metadata, and fused-MoE kernel execution) for NoDA and DA-MoE.}
    \label{fig:s5-logits-graph}
    
\end{figure}

Our distribution-aware selection runs on the critical path; copying the routing data to the host to choose a kernel would reintroduce the CPU--GPU synchronization that conditional CUDA Graphs are meant to remove.
Table~\ref{tab:s5-condcg} isolates this cost with a microbenchmark: a producer kernel computes a value that selects which of two follow-up kernels to launch, a data-dependent branch analogous to choosing a fused MoE kernel from the per-expert token counts.
Routing this branch on the host adds about 15~$\mu$s of overhead per dispatch over a branch-free fixed graph.
Keeping the branch on the GPU with a conditional CUDA graph cuts this to about 8~$\mu$s.
The per-graph overhead of our DA-MoE workflow, including histogram construction, sorting and exemplar matching, and the conditional branch, is about 14~$\mu$s.

Fig.~\ref{fig:s5-logits-graph} measures the end-to-end MoE-layer latency. The full graph includes router selection, routing metadata, GPU-resident DA-MoE dispatch, and fused-MoE kernel execution.
DA-MoE improves geomean end-to-end latency by 1.15$\times$ on DeepSeek-V3 and 1.26$\times$ on Kimi~K2, peaking at token count~128 (1.37$\times$ and 1.51$\times$).
These closely track the fused-MoE-only speedups of 1.16$\times$ and 1.29$\times$ from Fig.~\ref{fig:s5-primary-fc-bar}, confirming that the $\approx$14~$\mu$s dispatch overhead is small relative to MoE-layer latency and erodes the kernel-selection benefit only marginally.
Overall, conditional CUDA Graphs make GPU-resident distribution-aware dispatch practical by eliminating CPU-GPU synchronization.

\subsection{Distribution Modeling Accuracy}
\label{sec:eval-dist}

\begin{figure}[t]
    \centering
    \includegraphics[width=0.9\linewidth]{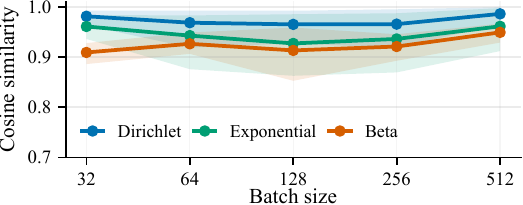}
    \caption{Cosine similarity between real sorted expert-count distributions (DeepSeek-V3, MMLU prefill tokens) and synthetic families matched to the same $eff\_exp^*$ value. Shaded regions show the 10th--90th percentile over sampled real distributions.}
    \label{fig:s5-dist-similarity}
    
\end{figure}

DA-MoE's offline auto-tuning is driven by synthetic routing distributions, so its validity rests on whether those distributions preserve the shape of real expert-count vectors once matched to the target $eff\_exp^*$ value.
Fig.~\ref{fig:s5-dist-similarity} compares three synthetic families---Beta, exponential, and Dirichlet---against real routing samples collected from DeepSeek-V3 on the prefill tokens of the MMLU dataset.
Each family is parameterized to match the sample's $eff\_exp^*$ value and then compared using cosine similarity over sorted, L2-normalized expert-count vectors.

The Dirichlet family gives the strongest match across token counts 32--512, with a geomean cosine similarity of 0.97, compared with 0.92 for Beta and 0.94 for exponential.
Beta requires a brute-force two-parameter search, for which we use a grid spacing of 0.1, whereas exponential and Dirichlet match the target $eff\_exp$ value through binary search.
This result confirms that a Dirichlet sweep is sufficient for offline tuning, as its mathematical formulation coincides with the expert-selection process, which is a distribution of distributions.

\subsection{Ablation Study: Routing Skew}
\label{sec:eval-ablation-imbalance}

\begin{table}[t]
\centering
\small
\caption{Fused-MoE latency speedup over NoDA (higher is better) for DeepSeek-V3
at EP=4 across imbalance factors.}
\label{tab:s5-ablation-imbalance}
\resizebox{0.95\linewidth}{!}{
\begin{tabular}{lcccc|c}
\toprule
\textbf{Imb ($f$)} & \textbf{TC 32} & \textbf{TC 64} & \textbf{TC 128} & \textbf{TC 256} & \textbf{Geomean} \\
\midrule
0   & 1.03$\times$ & 1.18$\times$ & 1.40$\times$ & 1.25$\times$ & 1.21$\times$ \\
10  & 1.06$\times$ & 1.35$\times$ & 1.56$\times$ & 1.31$\times$ & 1.31$\times$ \\
20  & 1.10$\times$ & 1.33$\times$ & 1.66$\times$ & 1.34$\times$ & 1.34$\times$ \\
50  & 1.35$\times$ & 1.68$\times$ & 1.95$\times$ & 1.43$\times$ & 1.59$\times$ \\
100 & 1.41$\times$ & 1.77$\times$ & 2.05$\times$ & 1.47$\times$ & 1.66$\times$ \\
\bottomrule
\end{tabular}
}
\end{table}

Autoregressive LLMs sample each next token from a stochastic probability distribution. Consequently, many evaluation protocols generate multiple responses for the same prompt to provide several attempts at a task. This experiment evaluates whether DA-MoE can effectively serve such workloads, in which repeated prompts create highly concentrated expert-routing distributions.

We introduce an imbalance factor $f \in \{0, 10, 20, 50, 100\}$. For $f>0$, we partition each batch into groups of approximately $f\%$ of its prompts and replace the prompts in each group with a shared prompt. The batch therefore contains $100/f$ distinct prompts, each repeated within its group, which concentrates the per-expert token counts without changing the nominal token count. At the two endpoints, $f=0$ leaves every prompt distinct, matching the unconstrained workload, whereas $f=100$ collapses the batch to a single repeated prompt and produces maximally skewed routing.

Table~\ref{tab:s5-ablation-imbalance} reports fused-MoE latency speedup over NoDA for DeepSeek-V3 at EP=4. The speedup grows monotonically with routing skew, increasing from a geomean of 1.21$\times$ at $f=0$ to 1.66$\times$ at $f=100$. These results show that DA-MoE is particularly effective for repeated-prompt workloads, where concentrated routing creates greater opportunity for distribution-aware kernel selection.

\subsection{Ablation Study: Expert-Parallelism Degree}
\label{sec:eval-ablation-ep}

\begin{table}[t]
\centering
\small
\caption{Fused-MoE latency speedup (NoDA latency / DA-MoE latency, higher is better) for DeepSeek-V3
across expert-parallelism degrees, at token counts 32--256 (imbalance factor 0). EP=1 corresponds to pure
pipeline parallelism (PP=4). Geomean is over token counts 32--256.}
\label{tab:s5-ablation-ep}
\begin{tabular}{lccccc}
\toprule
\textbf{EP} & \textbf{TC 32} & \textbf{TC 64} & \textbf{TC 128} & \textbf{TC 256} & \textbf{Geomean} \\
\midrule
1 (PP=4) & 1.00$\times$ & 1.00$\times$ & 1.01$\times$ & 1.00$\times$ & 1.00$\times$ \\
2        & 1.02$\times$ & 1.01$\times$ & 1.06$\times$ & 1.23$\times$ & 1.07$\times$ \\
4        & 1.03$\times$ & 1.18$\times$ & 1.40$\times$ & 1.25$\times$ & 1.21$\times$ \\
8        & 1.17$\times$ & 1.32$\times$ & 1.23$\times$ & 1.00$\times$ & 1.17$\times$ \\
\bottomrule
\end{tabular}
\end{table}

Expert parallelism sets how many devices share the expert pool, and a higher degree reduces the total routed tokens processed by each device's local experts, amplifying per-device routing imbalance.
We evaluate DeepSeek-V3 at EP $\in \{1,2,4,8\}$ with token counts 32--256.
EP=1 corresponds to pure pipeline parallelism with no cross-device expert sharding.

Table~\ref{tab:s5-ablation-ep} shows that DA-MoE delivers no measurable benefit at EP=1, with a geomean speedup of 1.0$\times$.
At EP=1, the fused kernel is selected once for the full 256-expert pool on each pipeline stage.
The local workload therefore combines many small expert GEMMs with different token counts, and one kernel configuration must serve this broad mixture rather than a narrow routing regime.
Under this aggregated view, changing the kernel for one hot subset provides limited benefit because the same choice is still applied to many other experts with different counts.
The static bucket choice is already a robust selection, leaving little room for DA-MoE to specialize.

As EP increases, each device sees a smaller shard of the expert pool, so local hot and cold patterns are less averaged out and the runtime selector can match a more specific routing regime.
The benefit grows to 1.07$\times$ at EP=2 and 1.21$\times$ at EP=4, then eases to 1.17$\times$ at EP=8.
Across EP=2--8, DA-MoE consistently outperforms the static baseline.

\section{Discussion \& Architectural Implications}
\label{sec:discussion}

DA-MoE shows that the live routing histogram is useful for choosing a faster MoE kernel, but this choice must be made after routing and before expert computation. Its cost lies directly on the decoding critical path. We draw two architectural implications from this result. First, GPUs need a low cost GPU-resident branching mechanism for dynamic kernel execution. Second, spatial accelerators need tile flexibility to match the observed expert distribution.

\subsection{GPU-Side Branching Should Be a Hardware Primitive}

Conditional CUDA Graphs are necessary because DA-MoE's kernel choice depends on a routing histogram produced on the GPU. Branching on the CPU would require copying this value back to the host and synchronizing the decoding path. Table~\ref{tab:s5-condcg} confirms this cost: CPU-side branching adds about 15~$\mu$s of overhead, while a conditional CUDA Graph reduces the branch overhead to about 8~$\mu$s.

This device-side branch is still not free as the full DA-MoE selection path adds about 14~$\mu$s per graph, including histogram construction, sorting, exemplar matching, and conditional dispatch. DeepSeek-V3 and Kimi~K2 amortize this cost because their MoE workloads are large enough for better kernel selection to dominate the overhead. Smaller MoE variants, such as Qwen3 models~\cite{yang2025qwen3technicalreport}, may not expose enough latency savings, so distribution-aware dispatch is beneficial only when the kernel-selection gain exceeds the dispatch cost.

A similar issue appears on AMD accelerators and their programming interface, HIP. HIP Graphs supports graph capture and replay, but do not expose an equivalent of CUDA's device-side conditional \texttt{SWITCH} node~\cite{hipgraph2026}. DA-MoE on HIP would therefore need CPU copy-back, removing the benefit of GPU-resident online dispatch.
Thus, graph replay alone is insufficient for MoE serving because the runtime also requires a low-overhead GPU-resident control flow.

\subsection{Spatial Accelerators Need Tile Flexibility}

The tile tradeoff is not specific to Tensor Cores. On systolic-array accelerators, the analogous choice is how much array area to allocate to each expert GEMM. A large tile improves reuse for hot experts, but wastes work when many experts receive few tokens. Section~\ref{sec:benchmark} shows that the preferred tile size changes with token count and routing skew.

ReSA~\cite{leeyeh2024resa} proposed a reconfigurable systolic array architecture that supports partitioning an array into smaller subarrays to handle multiple tiny GEMMs. For MoE workloads, these subarrays can be dynamically allocated to individual expert GEMMs using heterogeneous tile shapes. Our modeling framework directly applies to such reconfigurable spatial architectures, enabling quantitative performance analysis and optimal tile-shape selection under varying routing skews.

Overall, future inference accelerators need two forms of adaptivity: low-overhead device-side control flow, and tile-shape flexibility driven by the observed expert distribution.

\section{Related Work}
\label{sec:related}

\textbf{MoE inference systems and runtime dispatch.}
Production inference systems such as TensorRT~LLM, FlashInfer, and vLLM~\cite{trtllm,ye2025flashinfer,kwon2023vllm} support modern MoE models~\cite{switchtransformer,jiang2024mixtralexperts,deepseekai2025deepseekv32pushingfrontieropen,kimiteam2026kimik2openagentic} with fused MoE kernels, graph replay, and auto-tuning. Their dispatch policies are primarily keyed by static properties such as token-count buckets. DA-MoE adds the live routing distribution as a dispatch signal, selecting among existing kernel variants without changing the model or serving API. Some inference methods improve hardware utilization by changing expert choices~\cite{zeng2024topkmoe,yi2025edgemoe}, but they can incur accuracy tradeoffs.

\textbf{MoE training kernels and sparse expert computation.}
Training-oriented MoE systems reduce sparse expert-computation overhead through block-sparse execution, scalable scheduling, or IO- and tile-aware optimization~\cite{megablocks,yuan2025xmoe,guo2026sonicmoe}. These techniques address padding and scheduling costs in training workloads, while DA-MoE targets runtime kernel dispatch for low-latency inference.

\textbf{GEMM tiling and configurable accelerators.}
Prior architectural work on matrix multiplication shows that tile shape and resource partitioning should match the memory hierarchy and tensor shape. CAKE~\cite{kung2021cake} studies multi-level cache-aware blocking for CPU GEMMs, while ReSA~\cite{leeyeh2024resa} reconfigures systolic-array resources for small neural network tensors. Our work applies the same principle to MoE inference, where routing dynamically changes the per-expert GEMM sizes seen by fused GPU kernels.

\textbf{Communication and memory movement in MoE inference.}
Prior MoE inference systems optimize data movement, communication, memory capacity, or routing-pattern forecasting~\cite{li2023lina,cao2025moelightning,huang2026cpugpu,yu2026patternschaos}. Some systems coordinate expert placement and communication to reduce distributed all-to-all overheads, while others target memory-constrained serving by managing expert placement across GPU and host memory. Recent characterization work also studies routing and data-movement patterns to forecast communication behavior during large-scale MoE serving. These works reduce inter-device and memory-system overheads in distributed or memory-constrained settings.

\textbf{MoE routing characterization.}
Existing MoE analyses often rely on aggregate statistics or collected traces~\cite{megablocks,yu2026patternschaos}, which are difficult to sweep systematically in hardware studies. We introduce the \textit{effective experts} metric and a Dirichlet-based reverse-modeling procedure that converts real routing traces into a controllable distribution family. DA-MoE uses the same distribution-aware view at runtime by matching the observed per-expert histogram against offline-tuned exemplars.

\section{Conclusion}
\label{sec:conclusion}

This paper demonstrates that MoE inference performance depends not only on the token-count bucket, but also on the runtime routing distribution within each bucket. We introduce effective experts as a compact metric for expert-utilization skew and develop a Dirichlet reverse-modeling method for generating controllable distributions. This enables systematic hardware studies of MoE routing skew and shows that the best fused-MoE kernel changes with routing skew, token count, and expert-parallelism degree. These results indicate that static per-bucket kernel dispatch is insufficient for modern MoE serving.

We further present DA-MoE, a distribution-aware kernel-dispatch runtime building on top of the open-sourced kernel framework FlashInfer that selects near-optimal fused-MoE kernels. DA-MoE combines offline-tuned distribution exemplars with GPU-resident online matching to select the kernel. On HumanEval-X serving traces, DA-MoE improves geomean fused-MoE latency by 1.16$\times$ on DeepSeek-V3 and 1.29$\times$ on Kimi~K2, with peak speedups of 1.40$\times$ and 1.56$\times$, respectively. It also improves end-to-end MoE-layer latency by 1.15$\times$ and 1.26$\times$ geomean, peaking at 1.37$\times$ and 1.51$\times$, respectively. Overall, these results show that routing-aware dispatch and low-overhead device-side control flow are important mechanisms for efficient sparse expert computation in future MoE runtimes and accelerators.

\newpage
\bibliographystyle{IEEEtran}
\bibliography{refs}

\end{document}